\documentclass[smallextended]{svjour3}       
\smartqed  
\usepackage{xcolor}
\usepackage{graphicx}
\usepackage[round]{natbib}
\usepackage{aas-macros}
\usepackage{upgreek}
\usepackage{amsmath,amssymb}
\graphicspath{{./}}
\newcommand{\ARIEL}{\textit{Ariel}}

\begin{document}
	
\title{ArielRad: the \ARIEL\ Radiometric Model\thanks{This work has been supported by ASI grant n. 2018.22.HH.O.}
}

\author{Lorenzo V. Mugnai \and
		Enzo Pascale \and
		Billy Edwards \and 
		Andreas Papageorgiou \and
		Subhajit Sarkar}

\institute{L. V. Mugnai \and E.Pascale
\at
	Dipartimento di Fisica, La Sapienza Universit\`a di Roma, Piazzale Aldo Moro 2, 00185 Roma, Italy
	\email{lorenzo.mugnai@uniroma1.it}
	\and
	B. Edwards \at 
	Department of Physics and Astronomy, University College London, Gower Street, London, WC1E 6BT, UK
	\and
	A. Papageorgiou 
		\and
    S. Sarkar
    \at
	School of Physics and Astronomy, Cardiff University, Queens Buildings, The Parade, Cardiff, CF24 3AA, UK 
}

\date{Received: date / Accepted: date}

\maketitle

\begin{abstract}
	ArielRad, the \ARIEL\ radiometric model, is a simulator developed to address the challenges in optimising the space mission science payload and to demonstrate its compliance with the performance requirements. \ARIEL, the Atmospheric Remote-Sensing Infrared Exoplanet Large-survey, has been selected by ESA as the M4 mission in the Cosmic Vision programme and, during its 4 years primary operation, will provide the first unbiased spectroscopic survey of a large and diverse sample of transiting exoplanet atmospheres. To allow for an accurate study of the mission, ArielRad  uses a physically motivated noise model to estimate contributions arising from stationary processes, and includes margins for correlated and time-dependent noise sources. We show that the measurement uncertainties are dominated by the photon statistic, and that an observing programme with about 1000 exoplanetary targets can be completed during the primary mission lifetime.
\end{abstract}

\section{Introduction}
\label{Intro} 

In the past 20 years more than 4000 exoplanets have been detected using space and ground based surveys, and many more are expected to be discovered in the coming years thanks to space missions such as \textit{TESS} \citep{Ricker2016}, \textit{CHEOPS} \citep{Cessa2017}, \textit{PLATO} \citep{Rauer2014} and \textit{GAIA} \citep{GAIA2016}, and to ground instrumentation such as \textit{HARPS} \citep{Mayor2003}, \textit{HATnet} \citep{Bakos2018},  \textit{WASP} \citep{Pollacco2006},  \textit{KELT} \citep{Pepper2018}, \textit{OGLE} \citep{Udalski2015}, \textit{NGTS} \citep{Wheatley2013} and many others.

Planets have been found to be ubiquitous in our Galaxy, have been detected around almost every type of star and \citet{Cassan2012} infer that on average every star in our Galaxy hosts one planetary companion. The exoplanets detected thus-far show a diversity in their masses, sizes, orbits, and, presumably, physical and chemical conditions unseen among the planets in our own Solar System.

However, the essential nature of these exoplanets remains elusive. We have little idea whether the planet chemistry is linked to the formation environment or whether the type of host star drives the physics and chemistry of the planet’s birth, and evolution \citep{Tinetti2018}.

Atmospheric spectroscopy holds the key to unlock the mysteries of the chemical and physical conditions of these alien worlds as well as their formation and evolutionary histories. Multi-band photometry and spectroscopy of transiting exoplanets \citep{Seager2000} is currently one of the most effective observational techniques for revealing the chemistry and thermodynamics of exoplanet atmospheres~\citep{Charbonneau2005, Tinetti2007, Sing2016, Madhusudhan2012, Huitson2012,Kreidberg201}. Photometric and spectroscopic light-curves of transiting exoplanets provide a measurement of the transmission (transit) or emission (eclipse) spectrum of an exo-atmosphere, and can be used to reveal chemical constituents, as well as the pressure and temperature profile, using retrieval techniques originally developed for the study of the Earth and Solar System planets, and adapted to the new field of investigation \citep[e.g.]{Irwin2008, Line2013, Waldmann2015, Gandhi2017}.

Existing astronomical instrumentation has allowed to study spectroscopically a few tens of exoplanets, selected among those that are more easily observable based on  their sizes and temperatures, and over a limited spectral range \cite[e.g.][]{Sing2016, Tsiaras2018}. A significantly larger population study is now required in order to decipher the secrets of the exoplanets and their diversity.

The Atmospheric Remote-Sensing Infrared Exoplanet Large-survey, \ARIEL, has been selected by ESA as the next medium class mission of the Cosmic Vision programme to spectroscopically characterise the atmospheres of a large and diverse sample of exoplanets. \ARIEL\ will largely focus on warm and hot exoplanets, taking advantage from their well mixed atmospheres that show minimal condensation and sequestration of high atomic weight metals such as C, O, N, S, Si. The \ARIEL\ science payload uses a 1-m class telescope to feed a multi-band photometer and spectrometers covering the wavelength range from 0.5\,$\mu$m to 7.8\,$\mu$m, to sample both the peak thermal emission of the exoplanet atmospheres, and the spectral signature of all major atmospheric gases (e.g. H$_2$O, CO, CO$_2$, NH$_3$, CH$_4$, HCN, H$_2$S, TiO, VO) and condensed species.

The \ARIEL\ payload design is investigated using detailed simulations of the astrophysical detection, that take into account mission design parameters such as flight duration and sky availability, and payload and astrophysical uncertainties. Margins are used on each estimate to ensure all performance predictions are derived under reasonably conservative assumptions. 

In this work we describe ArielRad, the \ARIEL\ radiometric model simulator used to assess the payload science performance and to demonstrate its compliance with the science requirements. ArielRad is the third simulation tool developed to assess the mission performance and follows ExoSim \citep{Sarkar2020}, and the \ARIEL\ ESA Radiometric Model (AERM), developed by the Space Agency \citep{Puig2015} to support the flow down of science requirements to instrument requirements the radiometric model  during the \ARIEL\ phase/A study \citep{Tinetti2018}. 

ExoSim is a end-to-end, time-domain simulator of \ARIEL\ observations. It evaluates photometric and spectroscopic light curves implementing a detailed description of the instrument design, source of uncertainties, and systematics of instrument and astrophysical origin. As such, the simulated observations produced by ExoSim are similar to those expected with \ARIEL\ and require a full data reduction pipeline to detrend the observations and reconstruct the planet spectrum. ExoSim has been fully validated using noise modelling as well as real measurements obtained with Hubble Space Telescope. ExoSim simulations allow us to study effects of spatially and temporally correlated noise processes such as the photometric uncertainties arising from the jitter of the line of sight, or from the activity of the host star. However, ExoSim analyses are computationally intensive and it is currently impractical to conduct studies on more than a few targets, until a fully automated data reduction pipeline is developed and validated in the next phase of the project. AERM overcomes this limitation implementing a simplified approach based on a radiometric modelling of the detection and of the uncertainties. These simplifications mean it is capable of assessing the confidence limit on the detection of emission and transmission spectra of hundreds of exoplanet targets. AERM implements a noise model calibrated using ExoSim estimates and delivers compatible estimates on test targets. However, the noise model implemented in AERM is a two parameters model that falls short in capturing the complexity of the \ARIEL\ payload design. Consequently, AERM provides an overly pessimistic prediction on some targets which makes this simulator unsuitable in assessing instrument design solutions. 

ArielRad has been written to derive payload requirements from science requirements through detailed error budgeting, to validate the compliance of the payload design with the science requirements, and to optimise the payload design evaluating instrument design solutions over a proposed target list comprising about 1000 exoplanets. ArielRad overcomes the limitations of AERM by implementing a detailed payload model, similar to that used by ExoSim, capable of describing all major instrument components. ExoSim computing limitations are overcome in ArielRad by implementing radiometric estimates of the uncertainties of the detection. Noise contributions that need to be estimated in the time domain, such as the photometric noise arising from the jitter of the line-of-sight, are imported in ArielRad from pre-computed ExoSim estimates. ArielRad is used to create and maintain the top level payload performance error budgets, allowing a balanced allocation of resources across the payload.

In this work we describe ArielRad, the models implemented, their validation,  and provide examples of how ArielRad can be used to support the \ARIEL\ mission development, leaving to a future work a detailed assessment of the \ARIEL\ mission performance. 

\section{\ARIEL\ instrument design and observational strategy}
\label{sec:design}

The \ARIEL\ payload design is briefly described in this section with more details available in \citet{Tinetti2016, Tinetti2018, Pascale2018}, and in the \ARIEL\ Assessment Study Reports\footnote{https://arielmission.space/ariel-publications/.}. The telescope is an off-axis $0.63 m^2$ Cassegrain with an elliptical primary mirror, that provides diffraction limited performance at wavelengths longer than $3 \mu m$; there is no need for imaging capabilities and the telescope is cooled to less than $70 K$. A refocusing mechanism actuates the secondary mirror to correct possible misalignments, that can occur at launch and during thermalisation. 

The flux collected by the primary aperture feeds two separated instrument modules. A dichroic mirror splits the light into two beams, one at wavelengths shorter than $ 1.95 \mu m$ and the second at wavelengths between $1.95 \mu m$ and $7.8 \mu m$. The first beam is fed to an instrument module containing three photometers (VISPhot, $0.5 \mu m - 0.6 \mu m$; FGS-1, $0.6\mu m - 0.80 \mu m$; FGS-2, $0.80 \mu m -1.1 \mu m$) and a slit-less prism spectrometer (NIRSpec, $1.1 \mu m - 1.95 \mu m$) with spectral resolving power $> 15$. The two photometers, FGS1 and FGS2, operate as Fine Guidance Sensors (FGS), providing both photometric and pointing information for the attitude and orbital control system (AOCS). The second instrument module, fed by the longer wavelength beam, hosts the \ARIEL\ Infrared Spectrometer (AIRS), which consists of two prism-dispersed channels: Channel 0 (CH0)  covering the $1.95 \mu m - 3.9 \mu m$ band with a spectral resolving power larger than 100, and Channel 1 (CH1)  covering the $3.9 \mu m - 7.8 \mu m$ band with a spectral resolving power larger than 30. The spectrometers have field stops (slits) at an intermediate image plane, that are wider than the telescope Point Spread Function (PSF) and are used to limit the stray-light and the diffuse emission from reaching the focal plane.

During its four years primary mission, \ARIEL\ will observe $\sim1000$ planets. To maximise the scientific return, the mission implements a three-tier observational and data analysis approach, where three different sample spectra are analysed with optimised spectral resolutions, wavelength intervals and Signal-to-Noise ratios (SNR). A summary of this observational strategy is given in Tab. \ref{tab:tier} with the first tier (Tier 1) being a low spectral resolution reconnaissance survey of $\sim 1000$ planets to address science questions on a large population of targets, such as to the fraction of planets covered by clouds or of small planets that have retained a H/He atmosphere. The second tier, Tier 2, which consists of $\sim50 \%$ of the planets from Tier 1, are studied with a higher spectral resolution, merging the data collected in two juxtaposed spectral bins. 
Tier 2 analysis searches for potential correlations between atmospheric chemistry and basic parameters such as planetary size, density, temperature, stellar type and metallicity. It allows investigations of chemical abundances, cloud characterisation and elemental composition. Finally, $\sim10 \%$ of Tier 1 planets are re-observed multiple times in Tier 3, and the data are analysed using the full spectral resolution provided by the payload to gain detailed knowledge of the planetary chemistry dynamics and temporal variability of the exoplanet atmospheres. 
 
\begin{table}
	\caption{Summary of the science addressed in each tier.}
	
	\begin{tabular}{p{2cm}p{2.5cm}p{6cm}}
		\hline\noalign{\smallskip}
		{\bf \centering Tier Name} & {\bf \centering Observational Strategy} & {\bf \centering Science case}\\ 
		\noalign{\smallskip}\hline\noalign{\smallskip}
		Tier 1: Reconnaissance survey  &
		Low Spectral Resolution observation of $\sim 1000$ planets in the VIS and IR, with SNR $\sim 7$  & 
		\vspace{-5mm}
		\begin{itemize}
			\item What fraction of planets are covered by clouds?
			\item What fraction of small planets have still retained H/He?
			\item Classification through colour-colour diagrams
			\item Constraining/removing degeneracies in the interpretation of mass-radius diagrams
			\item Albedo, bulk temperature and energy balance for sub-sample
		\end{itemize} \\     
		
		Tier 2: Deep survey  &
		Higher Spectral Resolution observations of a sub-sample in the VIS-IR &
		\vspace{-5mm}
		\begin{itemize}
			\item Main atmospheric component for small planets
			\item Chemical abundances of trace gases
			\item Atmospheric thermal structure (vertical/horizontal)
			\item Cloud characterisation
			\item Elemental composition
		\end{itemize} \\
		
		Tier 3: Benchmark planets  &
		Very best planets, re-observed multiple times with all techniques &
		\vspace{-5mm}
		\begin{itemize}
			\item Very detailed knowledge of the planetary chemistry dynamics
			\item Weather, spatial and temporal variability
		\end{itemize}    \\
	\noalign{\smallskip}\hline
	\end{tabular}
	\label{tab:tier}
\end{table}

\section{The ArielRad simulator}
\label{sec:ArielRadSimulator}

ArielRad is a radiometric simulator of the \ARIEL\ payload, it is implemented in Python and it is maintained by the \ARIEL\ Consortium. The software package comes with an exhaustive documentation and the primary inputs are contained in a XML configuration file describing the payload, an XML configuration file describing the mission parameters, and a CSV or spreadsheet table containing a list of target exoplanetary systems with their parameters. An instrument independent version of the radiometric simulator called ExoRad is publicly available on GitHub\footnote{https://github.com/ExObsSim/ExoRad2-public}.

The simulator evaluates the payload science performance by estimating the expected experimental uncertainties on measured exoplanet atmospheric spectra in emission and transmission. Each simulation starts with the generation of the source signal from the star. The signal is then propagated through the instrument to the detector focal planes (assuming no field stars), accounting for the transmission of each optical component and the dispersion of the prism spectrometers.

\begin{figure*}
	\centering
	\includegraphics[width = 0.9\textwidth]{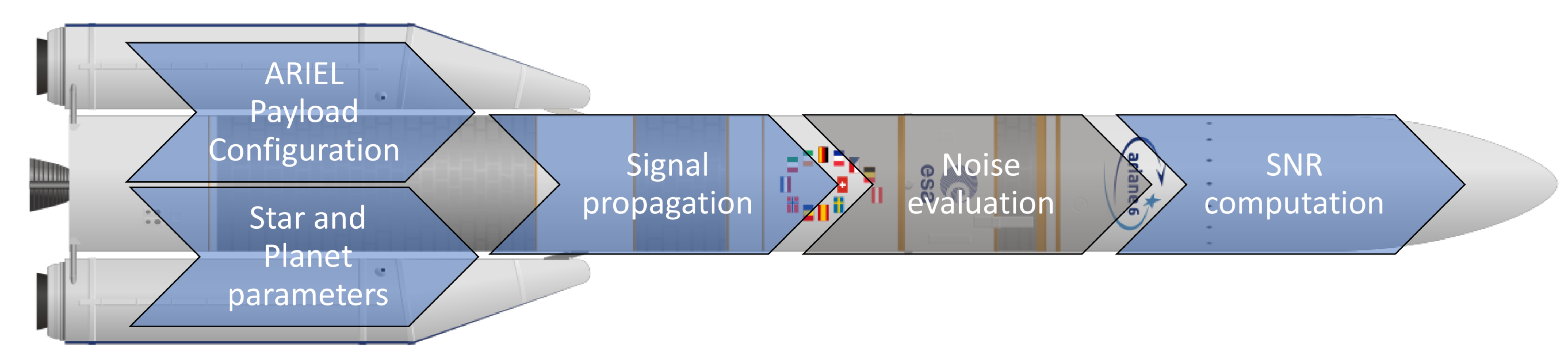}
	\caption{Simulator work flow. The simulation starts with two input files: a payload configuration file and a candidate planet list. ArielRad propagates the target host star signal thought the payload, then evaluates the noise. Finally, the simulator estimates the transit or eclipse observation and the resultant SNR.}
	\label{fig:simulator_scheme}
\end{figure*}

Uncertainty estimates account for detector noise (readout noise and gain noise, the latter arising from variations in acquisition electronic chain), dark current, photon noise from the star, the zodiacal background and includes instrument emission, and jitter noise. Margins are included to account for uncertainties in the current noise estimates and instrument performances. The simulation continues by estimating the planet atmospheric signal and thus the SNR of the detection when observing the target in transit or eclipse. Every candidate target observation is considered to last $2.5$ times the planet transit time, i.e. the time between the first and the last contact between the planetary and the stellar disks, $T_{14}$. This strategy allows the collection of data both in and out of transit for the light curve fit and the transit depth estimation. This parameter can be configured by the user as well as all other parameters described later.
The data is then binned according to the tier resolution, as will be done in \ARIEL\ data analysis. Tier 1 uses a low resolution spectroscopy (4 spectral resolution elements covering $1.1 -7.80 \mu m $); Tier 2 has spectral resolution $R \sim 10$ for $1.1<    \lambda <1.90 \mu m $, $R \sim 50$ for $1.95<   \lambda <3.90 \mu m $ and $R \sim 15$ for $3.90<    \lambda <7.80 \mu m $; Tier 3 uses the full $R$, that means $R=100,30$ for AIRS-CH0 and AIRS-CH1 respectively, and for NIRSpec, as the requirement is $R>15$, we use $R=20$. Finally, ArielRad estimates the number of observations required to achieve a desired SNR for each planet (e.g. SNR = 7) and from this the total observing time required on target. This process is summarised graphically in Fig. \ref{fig:simulator_scheme}) and each stage described in detail in the following sections.

\subsection{Source model}

ArielRad models the target star using a grid of synthetic Phoenix spectra from \citet{Baraffe2015}. Given a target, ArielRad selects the Phoenix spectrum that refers to a source with the most similar temperature, surface gravity and metallicity to the target star.
The Spectral Energy Density (SED) of the target star at the telescope input is evaluated as
\begin{equation}\label{eq:starSig}
S(\lambda_k) = \frac{1}{\lambda_k - \lambda_{k-1}}\frac{R_\star^2}{D^2}\int_{\lambda_{k-1}} ^{\lambda_{k}}S_{Ph}(\lambda)d\lambda
\end{equation} 
where $\lambda_k = \left(1+ \frac{1}{6000} \right)^k \lambda_{min}$ is a logarithmic spaced wavelength grid, determined starting from the sample spectral resolution of $6000$ and the minimum wavelength $\lambda_{min}$. This binning reduces the spectral resolution of the input spectra for computational efficiency, while preserving the total power. $R_\star$ is the target star radius, $D$ is the distance and $S_{ph}$ the input Phoenix spectrum. The units of $S(\lambda_k)$ are those of a spectral energy density.

\subsubsection{Diffuse radiation}
ArielRad includes contributions from the Zodiacal background, and instrument thermal emission.

The Zodiacal background emission is modelled using a modified version of the JWST-MIRI Zodiacal model \citep{Glasse2010}, scaled  according to the target position in the sky and the Zodi model of \citet{Kelsall1998}: 
\begin{equation}
I_{zodi}(\lambda) = A \cdot 3.5 10^{-14} \cdot BB(\lambda, 5500K) + B \cdot 3.58 10^{-8} \cdot BB(\lambda, 270K)
\end{equation}
where BB is the Planck law, and $A$ and $B$ are the fitted coefficients. Typically, $A$ and $B$ evaluate to $\sim 0.9$ at ecliptic poles, $\sim8$ at a Solar elongation angle of 55 deg and at an ecliptic latitude of 0 deg, and $\sim 2.5$ on average.  Therefore $A=B=2.5$ is assumed when target position coordinates are not available \citep{Pascale2015}. 

Instrument thermal emission, $I_{inst}$, is estimated by modelling each of the optical element as a Planck law at the element operating temperature and modified by a wavelength dependent emissivity. All the  reflective surfaces are made in aluminium, and we conservatively assume a 3\% emissivity. Same emissivity is assumed  for the refractive components.  All optical elements are passively cooled to 60K. Because of the almost isothermal \ARIEL\ design, the total instrument emission from optics is computed from the emission of one component, multiplied by the number of optical components along the light path. 

The exception is the contribution of the detector box, $I_{Inner}$, as the flux is not coming from the instrument field of view (FoV) but instead directly irradiates the pixels from all directions. This contribution is referred to as the ``Inner Sanctum" and it is estimated as the radiation emitted from a black body cavity at the detector operating temperature (i.e. unit emissivity is assumed).

\subsection{Payload configuration}
\begin{figure*}
	\centering
	\includegraphics[width=\textwidth]{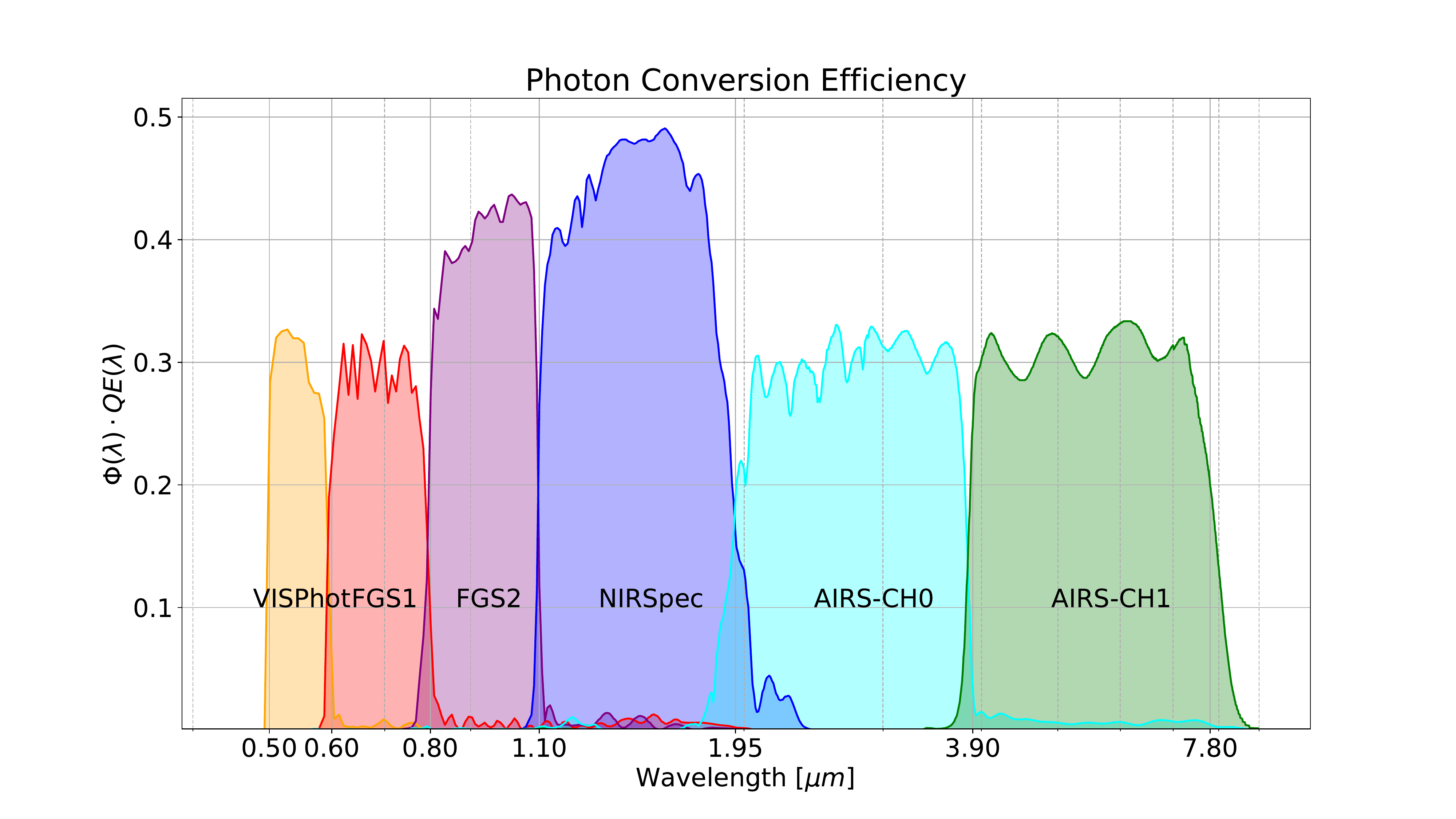}
	\caption{Total photon conversion efficiency, $\Phi_Y(\lambda) QE_Y(\lambda)$, as used in ArielRad.}
	\label{fig:throughput}
\end{figure*}

The payload is modelled using the current best estimates from the instrument system engineers. The transmission $\Phi_Y(\lambda)$, where $Y$ is one of VISPhot, FGS1, FGS2, NIRSpec, AIRS-CH0 or AIRS-CH1, is obtained by simulating the light path from the telescope to the detectors through the optics. The detector quantum efficiency $QE_Y(\lambda)$ is also dependent on  wavelength and defined for each channel. The photon conversion efficiency (PCE) is the product between transmission and quantum efficiency and it is shown in Fig. \ref{fig:throughput}. The lower PCE observed in VISPhot and FGS1 with respect to FGS2 and NIRSpec is caused by a lower detector QE at short wavelengths, while a similar PCE reduction at AIRS wavelengths is mainly a consequence of the refractive materials used in the mid-infrared.  The  Point Spread Functions (PSF) are estimated as a function of the wavelength using external software and included, allowing for wavelength interpolations. 

\subsection{Signal model}
\label{sec:signal_model}

The Point Response Function (PRF) is the PSF, normalised to unit volume and convolved with the pixel response, which is assumed to be a top-hat function, $H_{pix}(x,y)$. The effects of a non-flat pixel response function are only relevant for pointing jitter. As discussed later, this effect is estimated with ExoSim and imported in ArielRad as an additional noise contribution (see sec \ref{sec:jitter}). Hence,
\begin{equation}
PRF(x_l, y_m, \lambda) = \int \int PSF(x,y,\lambda)H_{pix}(x_l-x, y_m-y) dx dy
\end{equation}
where $l$ and $m$ are detector pixel indices, and  $\sum_{l,m} PRF(x_l, y_m,\lambda)=1$.

For each photometric channel, the signal of the target star $T$ is modelled as the sum of the signals sampled by all detector pixels within an elliptical aperture, the size of which depends on the PRF. Hence, this is estimated as
\begin{equation}
S_{T, phot}(\lambda_s) = A_{tel}\sum_{x_l, y_m \in A_p} PRF(x_l, y_m,\lambda_s) \int QE_Y(\lambda) \Phi_Y(\lambda) S(\lambda) \frac{\lambda}{hc} d \lambda 
\end{equation}
where $\lambda_s$ is the central wavelength of the photometric band (e.g. $0.55$, $0.70$ and $0.90 \mu m$ for VISPhot, FGS1 and FGS2 respectively). $A_{tel}$ is the telescope effective collecting area. $A_p$ is an elliptical aperture used for aperture photometry. Typical aperture sizes are chosen to encircle either 83.8\% or 91.0\%  of the energy in the PSF as ExoSim models indicate a reduction of jitter noise to negligible levels compared to other noise sources for these aperture sizes \citep{Sarkar2015}. The aperture, $A_p$, is defined over x,y pixels with coordinate indices $l,m$. 

The case of a spectroscopic channel is similar and the signal sampled by the detector is estimated as
\begin{equation}
S_{T, spec}(\lambda_s) = A_{tel} \sum_{x_l, y_m \in A_s} PRF(x_l, y_m) QE_Y(\lambda) \Phi_Y(\lambda) S(y_m) \frac{ \lambda}{hc}\Delta \lambda_m 
\end{equation}
where $A_s$ is now a rectangular box aperture of the size of the spectral bin along the y-axis, assumed to be parallel to the dispersion direction. Apertures are spaced such as to obtain the desired spectral binning $R$, which is 20, 100, 30 for NIRSpec, AIRS-CH0 and AIRS-CH1 respectively. $\Delta \lambda_m$ is the wavelength interval sampled by the pixel at coordinate $y_m$, and $\lambda_s$ is the wavelength at the centre of the spectral bin sampled by the spectrometer, defined as 
\begin{equation}\label{eq:lamb_s}
\lambda_s = \frac{1}{2} \left( \lambda_j + \lambda_{j+1} \right)
\end{equation}
where $\lambda_j$ is the wavelength at the bin edge defined by the recursive relation
\begin{equation}\label{eq:lamb_j}
\lambda_{j+1} = \lambda_j \left( 1+ \frac{1}{R} \right)
\end{equation}
and the focal plane wavelength solution maps pixel coordinates to wavelength ($\lambda_{j+1}, \lambda_{j} \rightarrow y_{i+1}, y_{i}$).

For the Zodiacal light contribution, the effect is proportional to the pixel solid angle,  $\Omega = \left( \frac{\Delta_{pix}}{f_{eff}} \right)^2$, where $\Delta_{pix}$ is the pixel size and $f_{eff}$ is the effective focal length. For the photometers and the slit-less NIRSpec spectrometer this contribution is given by
\begin{equation}\label{eq:zodi_phot}
S_{zodi,phot}(\lambda_s) = A_{tel} \Omega \int QE_Y(\lambda) \Phi_Y(\lambda) I_{zodi}(\lambda) \frac{\lambda}{hc} d \lambda 
\end{equation}

However, the AIRS instrument includes a field stop that being wider than the input PSF has no effects on a point source target, but acts as a slit for diffuse radiation. Therefore, Zodiacal light must be modelled differently for AIRS channels. ArielRad simulates the signal incoming to the detector as the convolution between the Zodiacal light and the field stop/slit. If the slit width expressed in number of pixels at the focal plane is $L$, and the spectral resolving power computed at a certain $\lambda_0$ is $R(\lambda_0)$, the detector receives diffuse radiation over the wavelength range $\left( \lambda_j - \frac{L \lambda_0}{4R(\lambda_0)} , \lambda_j + \frac{L \lambda_0}{4R(\lambda_0)} \right)$, and not over the full range of wavelengths accepted by the filter, so
\begin{equation} \label{eq:zodi_spec}
S_{zodi,spec}(\lambda_s) = A_{tel} \Omega \int_{\lambda_c - \frac{L \lambda_0}{4R(\lambda_0)}} ^{\lambda_c + \frac{L \lambda_0}{4R(\lambda_0)}} QE_Y(\lambda) \Phi_Y(\lambda) I_{zodi}(y_m) \frac{ \lambda}{hc} \Delta \lambda_m
\end{equation}
where $\lambda_0$ is $1.95$ and $3.90 \mu m$ and $R(\lambda_0)$ is $R(1.95) = 100$ and $R(3.90) = 30$ for AIRS-CH0 and AIRS-CH1 respectively.  

ArielRad models the diffuse light coming from instrument emission by substituting $I_{zodi}(\lambda)$ with $n \epsilon I_{instr}(\lambda)$ in Eq. \ref{eq:zodi_phot} and \ref{eq:zodi_spec}, where we are assuming that all the $n$ optical elements (lenses, mirrors, etc.) have the same emissivity, $\epsilon$. If the components are modelled with different emissivity,  $\sum_{i=1}^n \epsilon_i I_{instr}(\lambda)$ is used instead.

The Inner Sanctum contribution is the same for all  detector focal planes, as it originates from the emission of their enclosures, and it is proportional to the pixel surface:
\begin{equation}
S_{Inner}(\lambda_s) = \Delta_{pix}^2 \left( \pi - \Omega \right)  \int QE_Y(\lambda) I_{Inner}(\lambda) \frac{\lambda}{hc}d \lambda 
\end{equation}
Diffuse backgrounds add a DC offset in the measured stellar signal. ArielRad assumes that these are removed using standard techniques, e.g. aperture photometry, and only their (uncorrelated) contribution to the noise budget is considered later in this work. 

The resultant total signal in one spectral bin is the sum of all the previous contributions
\begin{equation}
S_Y(\lambda_s) = S_{T,Y}(\lambda_s) +  S_{zodi,Y}(\lambda_s) + S_{instr,Y}(\lambda_s) + S_{Inner}(\lambda_s)
\end{equation}
where in our notation $Y$ can be one of VISPhot, FGS1, FGS2, NIRSpec, AIRS-CH0 or AIRS-CH1. $S_{Y}(\lambda_s)$  has units of counts (electrons) per second.

\subsection{Noise model}\label{sec:noisemodel}

\begin{figure}
\centering
\includegraphics[width = 0.9\textwidth]{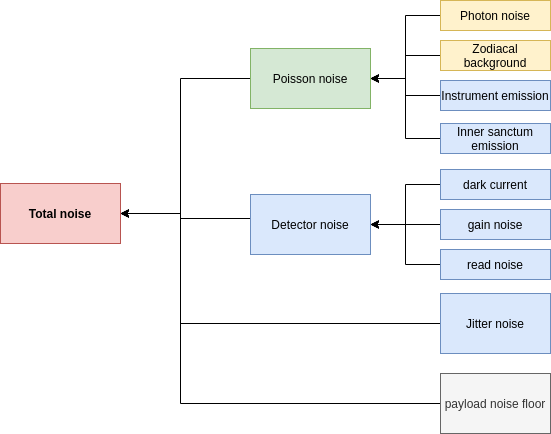}
\caption{Noise tree diagram listing all noise contributors considered in Eq. \ref{eq:noise}. The Poisson Noise term depends on signals collected by the telescope (in yellow the target and the Zodiacal background photon noise), and on signals of instrument origin (in blue, photon noise from instrument and inner sanctum self emission). The detector noise contributors are from dark current, gain and read noise. The Jitter noise term is discussed separately from the rest of the noise terms as it is imported as a model from ExoSim simulations. The payload noise floor prevents noise from integrating down indefinitely.}
\label{fig:noise_tree}
\end{figure}

ArielRad simulates the noise for each spectral bin from the signal estimate. In a real instrument, noise sources act at every stage of the detection chain and can be stationary and non-stationary. ArielRad estimates the contributions of noise components that are stationary random processes, such as Poisson noise and detector noise. It also includes Jitter noise (provided from an ExoSim simulation),  margins for other noise contributions, and  a noise floor. Fig. \ref{fig:noise_tree} shows a noise tree including all the noise sources considered.

The variance of each spectral bin is modelled as
\begin{equation}\label{eq:noise}
VAR[S_Y(\lambda_s)] = 
    \begin{cases} 
    (1+\chi_Y) VAR_P[S_Y(\lambda_s)] \,\frac{1}{\Delta t_{int}} + & \mbox{(Poisson noise term)} \\
    +VAR_D[S_Y(\lambda_s)]\,\frac{1}{\Delta t_{int}} + & \mbox{(Detector noise term)} \\
    +VAR_J[S_Y(\lambda_s)]\,\frac{1}{ \Delta t_{int}} + & \mbox{(Jitter noise term)} \\
    +\left[ p_0 S_{T,Y}(\lambda_s) \right]^2 & \mbox{(Payload noise floor)}
    \end{cases}
\end{equation}
where all quantities except $p_0$ are estimated at a 1\,hr integration time and at longer time scales are assumed to decrease as the inverse of $\Delta t_{int}$, the integration time in hours. Time scales longer than  1\,hr are expected to be above the correlation time scale of coloured noise processes considered. $\chi_Y$ is a margin added to include noise sources that cannot be simulated with ArielRad, such as correlated noise, time-dependent effects, and unknown unknowns. Typically, $\chi = 0.4$ is assumed, allowing for approximately $20$\% margin on top the photon noise that, from one side, it is conservatively larger than what has been achieved with \textit{Hubble}/WFC3 and \textit{Spitzer}/IRAC \citep[$\sim 15$\%,][]{Zellem2019}, but from the other it is adequate for the current stage of the mission design.

The \textit{Payload noise floor}, $p_0$, models low frequencies instabilities (e.g. Brownian or pink noise processes) that prevents noise to integrate down indefinitely with time. Following \citet{Greene2016}, we use $p_0 = 20$\,ppm of the incoming signal $S_{T,Y}(\lambda_s)$. See also \citet {Beichman2014,Molliere2017} for an example of this approach.

\subsubsection{Poisson Noise}
The Poisson noise term includes all the photon noise contributions: target source, zodiacal background, inner sanctum  and instrument emission. Hence,  the Poisson variance at 1\,hr time scale is
\begin{equation}
VAR_P[S_Y(\lambda_s)] = \frac{1\,{\rm hr}}{3600\,{\rm s}} \frac{1 }{\eta } \frac{6(n^2+1)}{5n(n+1)} S_{Y}(\lambda_s)
\label{eq:phot_noise_1}
\end{equation}
The factor $3600\,s$ normalises the variance at 1\,hr time scale and the parameter $\eta \lesssim 1$ is an efficiency accounting for detector reset events needed to prevent saturation and includes the temporal gap occurring between the end of the reset event and the first read. As the detector is read in $n$  consecutive non-destructive reads (NDRs) during each exposure, Eq. \ref{eq:phot_noise_1} is multiplied by  $\frac{6(n^2+1)}{5n(n+1)}$ as discussed in \citet{Rauscher2007} (MULTIACCUM readout strategy, see their Equation 1). This is appropriate as the number of group elements is 1 in the \ARIEL\ baseline payload design. The minimum value of $n$ is 2, i.e. correlated double sampling (CDS) required to remove the detector KTC noise;  $\frac{6(n^2+1)}{5n(n+1)} $ evaluates to 1 when $n=2$. NDRs are read at a constant cadence and $n$ depends from the detector saturation time
\begin{equation}
\label{eq:tsat}
T_{sat} = \frac{f_{WD} \cdot WD}{S_{Y, max}}
\end{equation}
where $WD$ is the detector well depth, $f_{WD}$ is the fraction of well depth used (e.g. $f_{WD} = 0.75$, \citealp{Berta2012}), and $S_{Y,max}$ is the largest pixel signal in the channel $Y$, that is estimated by ArielRad, but is not explicitly discussed here for the sake of conciseness. $VAR_P[S_Y(\lambda_s)]$ has units of s$^{-2}$hr. 

\subsubsection{Detector Noise}
The detector noise variance  can be divided in three terms, read noise, gain noise and dark current. 
The dark current signal depends on the detector pixel dark current, $I_{dark}$, and on the number of pixels, $N_{pix}$,  included in the photometric apertures $A_p$ or $A_s$  defined earlier,  with $S_{dark \, curr} = N_{pix}I_{dark}$. The detector noise variance at 1\,hr  is
\begin{equation}
\begin{aligned}
VAR_D[S_Y(\lambda_s)] & =  \frac{1\,{\rm hr}}{3600\,{\rm s}} \frac{12(n-1)}{n(n+1)} \frac{N_{pix}\sigma_{rd,Y}^2}{\eta^2 T_{sat}} + \\
& + \sigma^2_{gain, Y} S_Y^2(\lambda_s) + \\
& + \frac{1\,{\rm hr}}{3600\,{\rm s}} \frac{1 }{\eta } \frac{6(n^2+1)}{5n(n+1)} S_{dark\, curr} 
\label{eq:photon_noise}
\end{aligned}
\end{equation}
where $\sigma_{rd, Y}^2$ is the noise variance on each individual NDR and has no units. Following \citet{Rauscher2007}, the factor  $\frac{12(n-1)}{n(n+1)}$ accounts for a line fit to the NDR ramp in one exposure, and decreases as $n$ increases.  

The gain noise , $ \sigma_{gain, Y}$ in units of $\sqrt{\rm hr}$,  accounts for instabilities of the electronic acquisition chain (amplifiers, digitisers, etc.), assumed post-processing, i.e. after common modes are removed using housekeeping information. With $T_{sat}$ in units of seconds, $VAR_D[S_Y(\lambda_s)]$ has units of s$^{-2}$hr. 


\subsubsection{Jitter Noise} 
\label{sec:jitter}

Pointing drifts and jitter of the line-of-sight (LoS) manifest themselves in the observed data product via two mechanisms: 1) the drifting of the spectrum along the spectral dispersion axis of the detector; 2) the drift of the spectrum along the cross-dispersion (spatial) direction. The effect of jitter on the observed time series is the introduction of noise, characterised by the power-spectrum of the telescope pointing (usually not stationary).  It is correlated in time, as the power-spectrum is not constant in frequency. The amplitude of the resultant photometric scatter depends on the amount of spectral/spatial displacement of the spectrum, the monochromatic PSF of the instruments, the detector pixel response function (intra-pixel response) and the amplitude of the inter-pixel variations (i.e. QE variations across the focal plane detector array). Modelling the complexity of the jitter noise effect is beyond the capabilities of a radiometric model, therefore ArielRad imports jitter noise models from ExoSim simulations that provides variance vs wavelength at a timescale of 1\,hr, i.e.  $VAR_J([S_Y(\lambda_s)]$ in Eq.~\ref{eq:noise}. At longer timescales, jitter noise is to good approximation time uncorrelated and can be therefore scaled at any desired observing time longer than 1\,hr. 

\subsubsection{Additional noise contributions}
As ArielRad is a radiometric model of the \ARIEL\ payload performance, it cannot simulate non-stationary noise processes, or processes that require a more sophisticated simulation strategy, i.e. time-domain. One example of the latter is the already discussed pointing jitter, that is accounted for in ArielRad using external modelling with ExoSim.

Detector persistence is an additional potential systematic that cannot be modelled with ArielRad, but {\em HST} observations have shown that it can be effectively corrected in data analysis \citep{Zhou2017}. Further, this effect is expected to be negligible in \ARIEL\ observations compared to {\em HST} as \ARIEL\ stares continuously at a target from an L2 orbit ensuring that detectors reach a steady state in a relatively short amount of time (few minutes at most).  

Detector non-linear behaviour is well characterised in detector testing activities  before launch and during commissioning. Coupled with a stable line-of-sight provided by the AOCS at the sub-pixel level, detectors pixels sample the same optical power level during an observation, therefore dwarfing, relatively to other sources of experimental uncertainties, this type of systematic that is also known to be amendable \citep{Jong2006}.

Throughput variations can be caused by thermoelastic deformations that can also induce temporal variations in the shape of the PSF. To minimise this effect, the \ARIEL\ payload design uses aluminium structures, including the telescope mirrors. Along with the thermal stability enjoyed by spacecraft in L2 orbits, this prevents driving any significant thermal gradient on the payload structure and subsystems. It is therefore expected by design that thermoelastic throughput variations during \ARIEL\ observations will be significantly below any detection limit.

\begin{figure*}
	\centering
	\includegraphics[width=\textwidth]{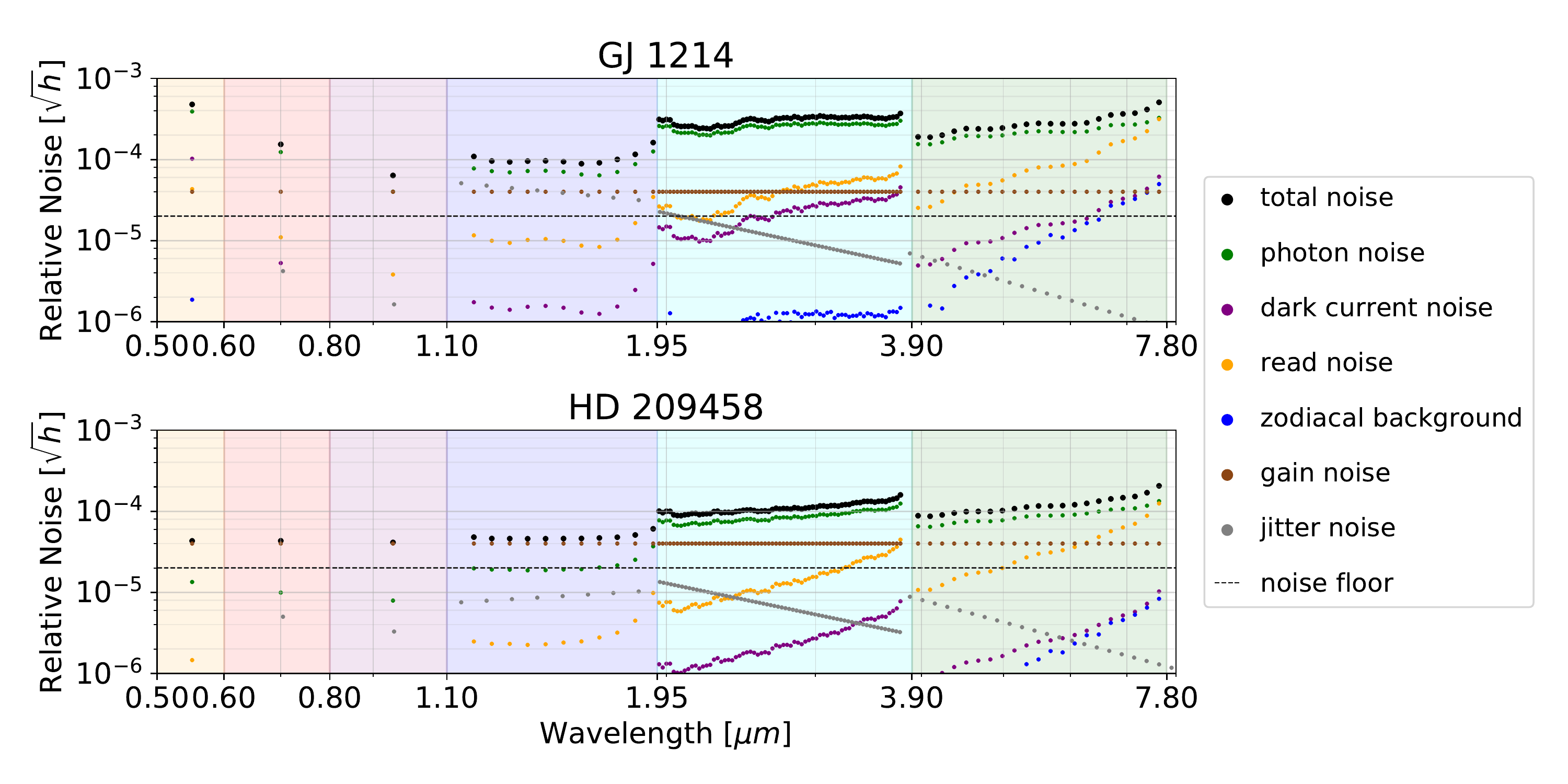}
	
	\caption{ArielRad noise budget example of the  \ARIEL\ payload at 1\,hr integration time, using Tier 3 binning. Two targets are considered: GJ 1214, a faint target for the \ARIEL\ mission, and HD 209458, a typical bright target. The total noise estimate is in black and includes a  20\% margin in excess of the target photon noise (green). Also shown are the detector dark current and read noise (purple and yellow, respectively),  photon noise from the zodiacal background (blue), gain noise (brown), and pointing jitter noise (gray). A noise floor is indicated by the horizontal dashed line.  The pointing jitter noise model is estimated using  ExoSim \citep{Sarkar2015} simulations with identical payload parametrization. Gain noise is assumed to be equal to 40\,ppm $\sqrt{\rm hr}$.  Photon noise from instrument and inner sanctum emission are omitted because are negligibly small and out of the scale of the diagram. . The \ARIEL\ payload model used in this budget is that at the mid of phase B1, it includes a noise floor of 20ppm and identical detector median dark current noise across all channels, that might result in an overestimate of up to a factor of 3 in all channels, but CH1.}
	
	\label{fig:noiseBudget}
\end{figure*}

\subsubsection{Output format}

ArielRad outputs the noise integrated over 1\,hr of observation relative to the target signal:
\begin{equation}
\sigma_{Y,1hr}(\lambda_s) = \frac{\sqrt{(1+\chi_Y)VAR_P + VAR_D + VAR_J}}{S_{T, Y}(\lambda_s)}    
\end{equation}
The units are hr$^{1/2}$. The relative noise achieved during one observation is
\begin{equation} \label{eq:noise1hr}
\sigma_{Y, \Delta t}^2(\lambda_s) = \frac{\sigma_{Y,1hr}^2(\lambda_s)}{\Delta t_{int}} + p_0^2 
\end{equation}where $\Delta t_{int}$ is the observing time in hours, and the payload noise floor $p_0$ is added in quadrature to prevent the noise to integrate down indefinitely as discussed earlier.

The payload (relative) noise budget is shown in Fig.~\ref{fig:noiseBudget} for a typical bright target star (HD 209458) and for a faint target star (GJ 1214). These stars are representative of respectively the bright and faint flux limits of the payload design requirements. The noise budget has been estimated at 1\,hr integration assuming that detectors are analysed in CDS samples, i.e.  $n=2$. The budget is sampled at Tier 3 resolution and it allows comparison of noise contributions to the total noise in any individual \ARIEL\ channel. A comparison of sensitivity among channels for any given target needs to take into account the different spectral binning used that is higher in CH0 ($R$ = 100) compared to NIRSpec ($R=20$) and CH1 ($R=30$).  A strong increase with wavelength is seen in the read and dark current noise components as a consequence of a decreasing stellar SED toward the red. 

As it is evident from the noise budget, the \ARIEL\ payload design reaches photon noise limited performances, allowing observations of even dimmer targets.

\subsubsection{Signal-to-Noise ratio}
In a transit or eclipse observation, the observable is the wavelength-dependent contrast-ratio that is the difference between the flux incoming from the extrasolar system star plus planet when the planet is moving in front (transit) or behind (secondary transit, i.e. eclipse) the star and when it is not:
\begin{equation}
CR(\lambda) = \frac{S_{OOT}(\lambda)-S_{IT}(\lambda)}{S_{OOT}(\lambda)}
\label{eq:cr}
\end{equation}
where the labels {\em OOT} and {\em IT} identify signals measured respectively out of transit and in transit.

For a primary transit the contrast ratio is evaluated from the comparison between the SED of the star $S_\star(\lambda)$ plus that of the planet $S_p(\lambda)$, $S_{OOT}(\lambda) = S_\star(\lambda) + S_p(\lambda)$, and the SED measured during the transit, that up to the first order depends on the radii ratio \citep{Mandel2002, Seager2002}. 
ArielRad simulates the contrast ratio in each spectral bin for the primary transit, considering also the portion of the star light passing through the planet atmosphere, as described in \citet{Seager2000} and \citet{Brown2001}: 
\begin{equation}
CR_{tot}(\lambda) = \left(\frac{R_p}{R_\star}\right)^2 + 2 \Delta z(\lambda) \frac{R_p}{R_\star^2}
\end{equation}
where $R_p$ and $R_\star$ are the planet and star radii respectively and $\Delta z(\lambda)$ is the atmospheric height. Because the interest is on the detection of the exoplant atmosphere, the ``signal'' is just the rightmost quantity \citep{Louie2018, Zellem2019} in the expression above, i.e.
\begin{equation}
CR(\lambda) =  2 \Delta z(\lambda) \frac{R_p}{R_\star^2}
\end{equation}
The atmospheric height is proportional to the scale height $H = k_B T/ \mu g$ where $k_B$ is the Boltzmann constant, $T$ is the temperature of the atmosphere, $\mu$ is the mean molecular weight and $g$ is the gravitational acceleration. The wavelength dependent constant of proportionality can be provided by atmospheric models or, following \citet{Tinetti2013}, set to 5 independently from the wavelength for a simple, yet representative performance estimate.

For the eclipse case, the contrast ratio is simply $CR(\lambda) = {S_p(\lambda)}/{S_\star(\lambda)}$ \citep{Charbonneau2005, Deming2005}, and the planetary SED is part due to the planet thermal emission, $S_{em}(\lambda)$, and part due to the reflected star light, $A_l(\lambda)$. Thus,

\begin{equation}
CR(\lambda) = \frac{S_{em}(\lambda) + A_l(\lambda)}{S_\star(\lambda)}
\end{equation}
The simulator estimates the contrast ratio in each spectral bin modelling the planet emission as a Black Body at the planet temperature, and computes the reflected light component according to \citet{Charbonneau_1999} as $A_l (\lambda)= \alpha(\lambda) \left( \frac{R_p}{a} \right)^2 S_\star(\lambda) $ where $\alpha(\lambda)$ is the geometric albedo and $a$ is the semi-major orbital axis.

ArielRad can estimate these contrast-ratios considering the observation of $S_{IT}$ lasting the time between the first and last contact, $T_{14}$, and $S_{OOT}$ lasting $\gamma \, T_{14}$, with $\gamma = 1.5$, topically, from current \ARIEL\ science requirements.
Using Eq. \ref{eq:noise1hr} and \ref{eq:cr}, the noise variance estimate on a contrast ratio measurement is
\begin{equation}
VAR(CR_Y, \lambda_s)=\sigma^2_{Y,1hr}(\lambda_s) \left[ 1 +\frac{1}{\gamma} \right] \frac{1}{T_{14}} + p_0^2
\end{equation}

and the Signal-to-Noise Ratio in each spectral bin is
\begin{equation}
SNR_Y(\lambda_s) = \frac{CR_Y(\lambda_s)}{\sqrt{VAR(CR_Y, \lambda_s)}}
\label{eq:snr}
\end{equation}
where we can substitute $CR(\lambda_s)$ with the estimated contrast ratio for transit or eclipse observations. The label $Y$ indicates that quantities are integrated over the photometric band or spectral bin of interest. Therefore ArielRad estimates three sets of SNR, one for each of the three \ARIEL\ tier discussed in Section \ref{sec:design}.

The SNR achieved in multiple, $N_{obs}$ observations of the same target extrasolar system is assumed to scale as $\sqrt{N_{obs}}$ , under the assumption that disturbances originate from stochastic processes that are uncorrelated over the time scale separating two observations, and longer. If this were not the case, it would be more appropriate to assume that the noise floor is not reduced by averaging multiple observations. However, there is currently no evidence supporting this as \textit{Hubble}/WFC3 and \textit{Spitzer}/IRAC observations have not yet reached a noise floor \citep{Zellem2019} .


\section{Validation \label{sec:validation}}
ArielRad has been validated by comparing its estimates to those of ExoSim and AERM introduced in Section \ref{Intro}. 

As ExoSim has been extensively validated against real astrophysical observations \citep{Sarkar2020}, the comparison between ArielRad and ExoSim estimates are the most interesting to investigate. For this, we chose to compare  the predictions made of $T_{sat}$ (Eq. \ref{eq:tsat}). A consistent result between EsoSim and ArielRad on this parameter implies that ArielRad captures the complexity of the payload design as thoroughly as ExoSim does, and validates the implementation of the radiometric algorithms. The comparison is therefore done implementing the same baseline \ARIEL\ model in the two simulators, and $T_{sat}$ is evaluated for the three target stars GJ~1214 (M4.5, mag$_K \simeq 8.8$), HD~209458 (G0V, mag$_K \simeq 6.3$) and HD~219134 (K3V, mag$_K \simeq 3.3$) that cover a range in brightness and temperature representative of potential \ARIEL\ targets \cite[e.g.][]{Edwards_2019}. The comparison is given in Table~\ref{tab:tsat} For the visible photometer and the infrared spectrometer;  it is found that the two models agree to better than 5\% on all targets.

\begin{table}[h!]
	\caption{Comparison between $T_{sat}$ estimates with ExoSim and ArielRad.}
	\centering
	\begin{tabular}{|c | c c c|} \hline
		~ & GJ 1214  & HD 209458 & HD 219134 \\
		Channel     & \multicolumn{3}{c|}{Percent variation}\\ \hline
		VISPhot     & -0.8 & -0.5 & -0.4 \\ 
		AIRS-CH0 & -2.9 & 1.0 & 1.2 \\ 
		AIRS-CH1 & 4.4 & 2.5 & 2.8 \\ 
		\hline
	\end{tabular}
	\label{tab:tsat}
\end{table}

\begin{figure}
	\centering
	\includegraphics[width = 0.45\textwidth]{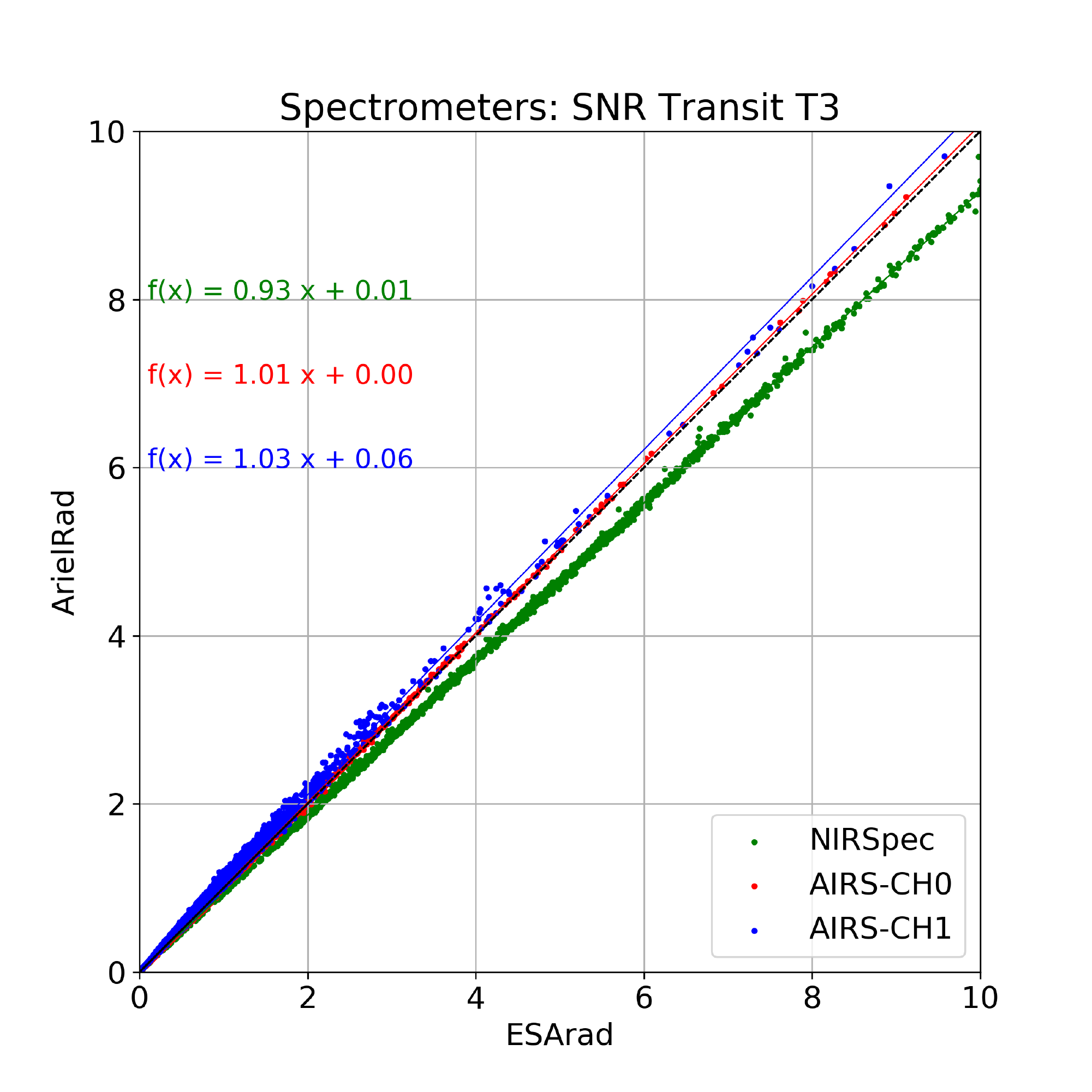}
	\includegraphics[width = 0.45\textwidth]{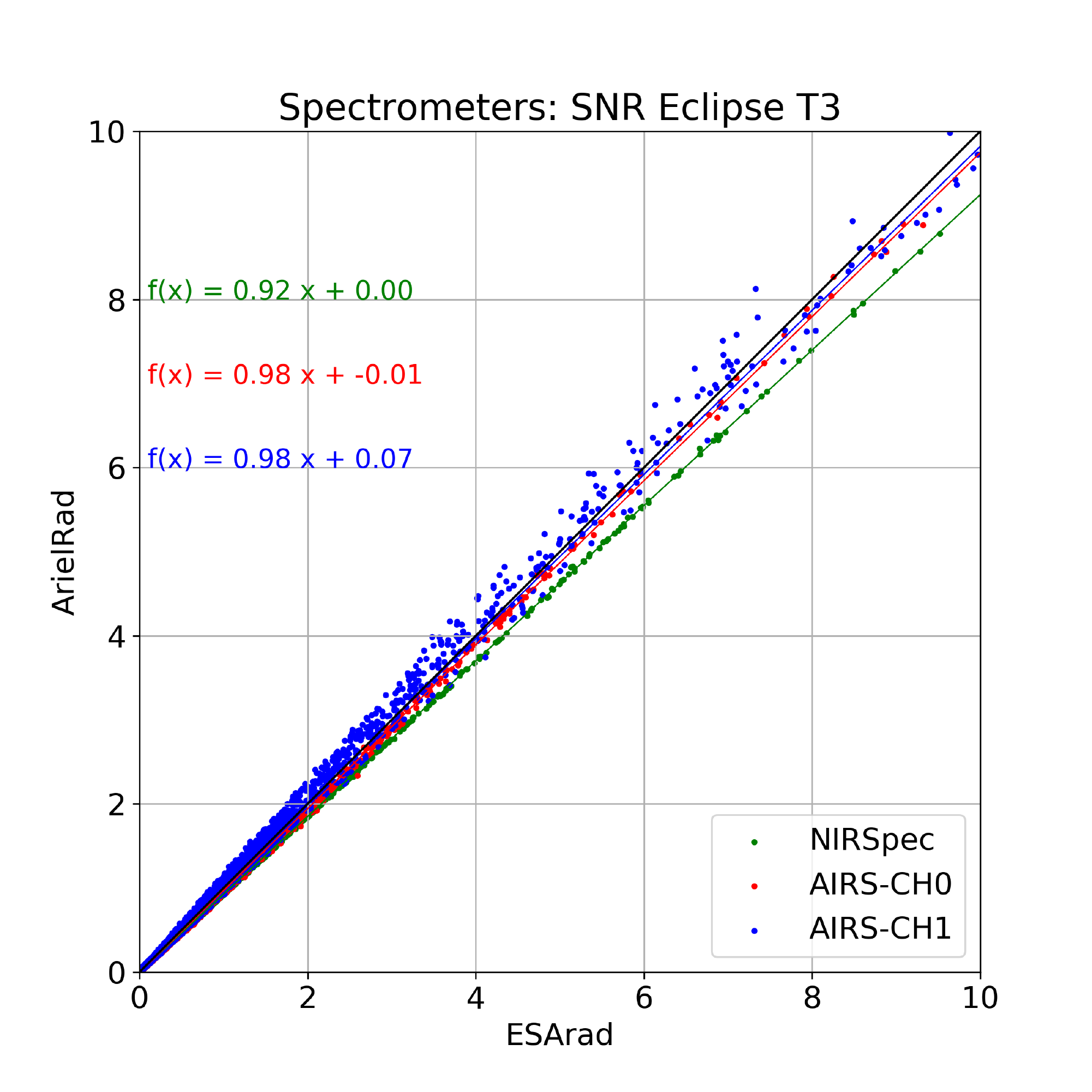}
	\caption{Comparison between ArielRad and AERM SNR estimates in one transit (\textit{left}) and one eclipse (\textit{right}) observation of a sample of 2500 candidate exoplanet atmospheres. Each datapoint is the average SNR across the NIRSpec (green), AIRC-CH0 (red) and AIRS-CH1 (blue) bands, binned at the \ARIEL\ tier 3 spectral resolution. Trend lines are shown with slope and intercept given in the text annotation in each panel. A black dotted line with unity slope is shown. While estimates for CH0 and CH1 are in good agreement among both models, AERM provides SNR estimates in NIRSpec that systematically and unrealistically larger than ArielRad. This is due to a limitation in the AERM algorithms implemented and it is further discussed in Section \ref{sec:validation}. }
	
	\label{fig:ArielRadvsESArad}
\end{figure}
The comparison between ArielRad and AERM provides validation of the SNR calculations summarised by Eq. \ref{eq:snr}, that means a validation of all calculations implemented concerning the estimate of the exoplanet atmospheric signatures observable during a transit or an eclipse and uncertainties. \citet{Puig2015}, and reference therein, detail the algorithms implemented by AERM, that estimates the noise during 1\,s of integration  as
\begin{equation}
\sigma_{Y,Total} = \sqrt{[S_{T, Y}(\lambda_s)+ S_{zodi, Y}(\lambda_s)] \cdot (1+\chi_Y) + N_{min,Y}(\lambda_s)}
\end{equation}The main difference with the ArielRad implementations is the $N_{min}(\lambda_s)$ term that combines the noise variance in a spectral bin or photometric band from detector dark current and instrument emissions. For the photometric channels, $N_{min, Y}(\lambda_s)$ is set to 400 s$^{-1}$. For the prism spectrometers, $N_{min,Y}(\lambda_s)\propto \lambda_s^3$, and it is set to 17\,s$^{-1}\mu$m$^{-3}$,20\,s$^{-1}\mu$m$^{-3}$, and 5\,s$^{-1}\mu$m$^{-3}$ at the blue end of NIRSpec, AIRS-CH0 and AIRS-CH1, respectively.  There is no provision for other noise sources in AERM, including detector readout and gain noise, nor pointing jitter. These effects are accounted for in the term $\chi_Y$ that is set to 0.2 for the photometers and NIRSpec, and to 0.3 for both AIRS channels. AERM further assume a photon conversion efficiency that, in each photometric or spectroscopic channel is wavelength independent. For this validation exercise, the ArielRad input configuration is adapted to match that in AERM, in terms of photon conversion efficiency, and matching dark currents to the equivalent  $N_{min,Y}
(\lambda_s)$. Furthermore, the parameters in Eq. \ref{eq:phot_noise_1} are chosen such that $n=2$ (i.e. AERM assumes CDS), $\eta = 1$ (i.e. AERM assumes 100\% sampling efficiency) and the photon noise margin parameter $\chi$ are set to those in use in AERM. $VAR_D$, $VAR_J$, and $p_0$ are set to zero in Eq. \ref{eq:noise}. With this, AERM and ArielRad implement effectively the same instrument model, and are run to evaluate the SNR achieved on one transit or one eclipse over  2500 candidate \ARIEL\ targets \citep{Edwards_2019}. Figure \ref{fig:ArielRadvsESArad} shows the comparison of the two model estimates of the average SNR over spectral bins in each spectrometer. We find that, on average, the SNR estimates agree to better than 2 or 3\% for AIRS channels.  For NIRSpec we find that on average ArielRad predicts SNR that are up to 8\% smaller than AERM's predictions. A further investigation has revealed the root cause of this discrepancy in the way AERM estimates the signals: while ArielRad integrates SEDs over wavelengths in a spectral bin, AERM uses the SED estimated at the blue end of each spectral bin. This always implies that signal in AERM are systematically larger resulting is smaller uncertainties. The effect is more evident in NIRSpec because the larger spectral bin width (R = 10, as impleneted in ESArad) compared to AIRS channels (R=100 and R = 30 in CH0 and CH1, respectively). 

\section{Use of ArielRad}
ArielRad has been developed to support the \ARIEL\ phase B study, leading to Mission Adoption by ESA in the Autumn 2020. ArielRad allows to evaluate the payload science performance over a large target list of thousands of potential exoplanetary targets and to assess the compliance of the payload design with the science requirements briefly discussed in Section \ref{sec:design}, with a more detailed discussion in \citet{Tinetti2018} and in the \ARIEL\ yellow book\footnote{https://sci.esa.int/s/8zPrb9w.}. ArielRad is the main tool used to assess the \ARIEL\ payload design solutions and provides a guide to optimise the payload to achieve requirements, and to maximise the \ARIEL\ science return beyond requirements, when possible.  

The observed diversity of exoplanets can only be investigate by surveying a large parameter space in planetary radii and masses, thermodynamic conditions, chemical properties and host star types. \ARIEL\ is designed to provide the first large survey of the atmospheres of about 1000 diverse planets and ArielRad is the tool used  to  craft  a target list that is compliant with this science mandate. \citet{Edwards_2019} used ArielRad simulations to provide a preliminary  mission reference sample (MRS) of 1000 planets. While the MRS is expected to evolve during the next phases of the project until launch in 2028, the ArielRad performance analysis demonstrates that the atmospheres of planets in the MRS can be characterised with a $SNR > 7$ during the 4 year nominal mission lifetime.  

\begin{figure*}[t]
	\centering
	\includegraphics[width=0.95\textwidth]{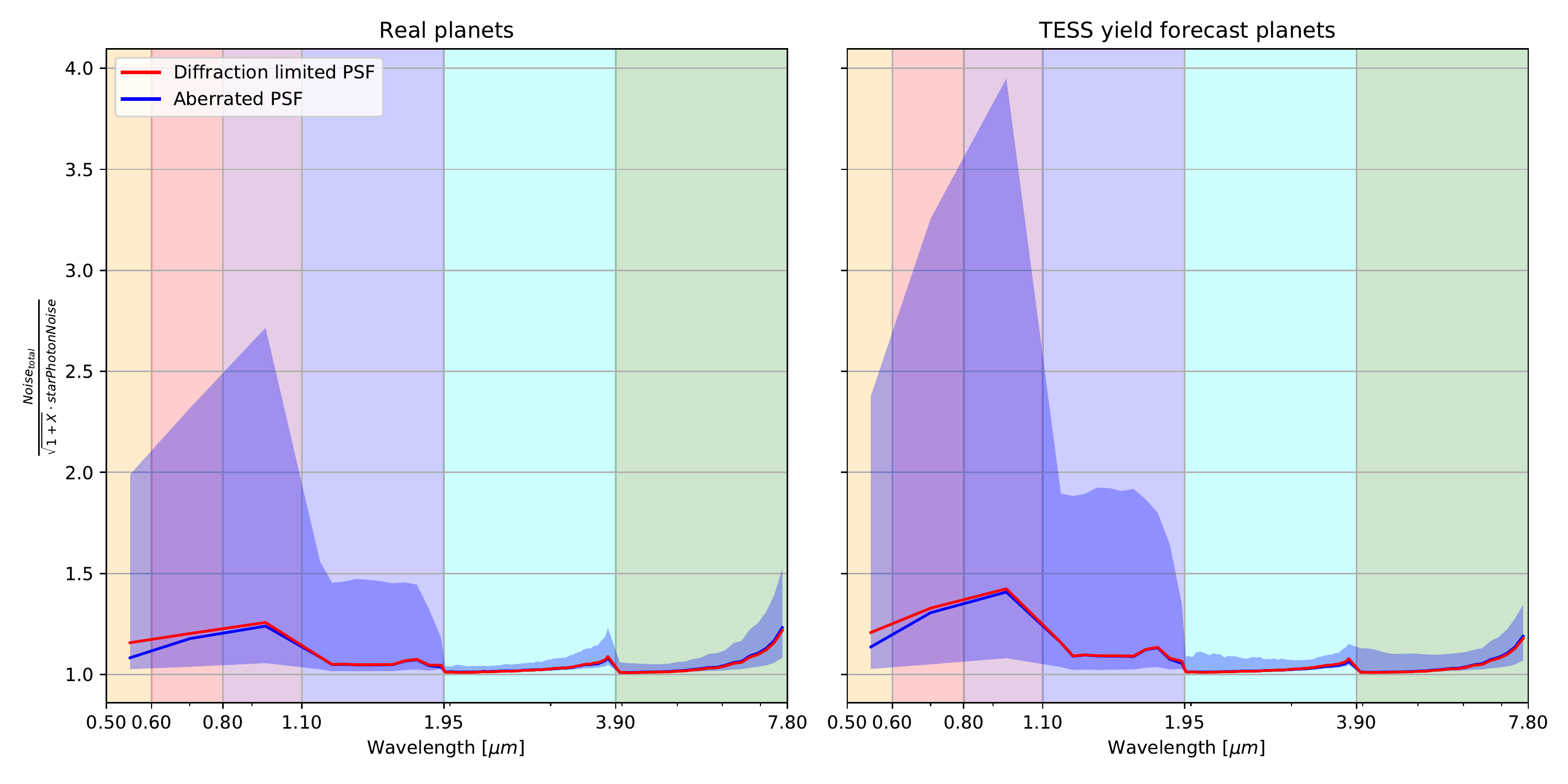} 
	\caption{Total noise to photon noise ratio in one hour integration for all \ARIEL\ MRS targets. Solid and dashed lines mark median values across all targets evaluated for diffraction limited (red) and aberrated (blue) optics. Blue area represents the dispersion of $95\%$ of simulated planets for the aberrated optics configuration. As the MRS contains exoplanets already discovered as well as expected TESS yields, these are separated respectively in the left and right panels. As discussed in the text, the apparent excess noise above photon noise at visible-nearIR wavelengths is due to the presence of cold M-type stellar targets, that are more numerous among TESS targets, and the assumed gain noise contribution. For these type of targets, observations in the nearIR-midIR are more sensitive showing the power of \ARIEL's IR bands that are always shot noise limited. }
	\label{fig:photonLimit}
\end{figure*}

As discussed in Section \ref{sec:noisemodel}, ArielRad provides a detailed description of the \ARIEL\ noise budget (Figure \ref{fig:noiseBudget}) on individual targets. The analysis can be extended to provide a comprehensive description of the payload performance over all targets in the MRS to show how \ARIEL\ achieves a photon noise limited performance on all targets, as shown in Figure~\ref{fig:photonLimit}.  The MRS of \citet{Edwards_2019} is used. It lists both exoplanets already discovered and expected TESS yields. At AIRS wavelengths ($\lambda_s > 1.95\,\mu$m) photon noise is the dominant source of uncertainty, dwarfing all other noise contributions. At shorter wavelength, photon noise is less important for a small, but significant number of targets, with detector gain noise playing a larger role in the noise budget. These are largely TESS targets around M-type (cold) host stars, demonstrating the power of the \ARIEL\ IR bands for these type of targets.The \ARIEL\ telescope is diffraction limited at a wavelength of 3\,$\mu$m and significant optical aberrations degrade the image quality at shorter wavelengths. However, \ARIEL\ works as a light bucket, and image quality is not relevant to achieve its performance requirements. This aspect is also investigated in Figure \ref{fig:photonLimit} where the analysis is done for both a diffraction limited instrument and using estimates of aberrated PSF from engineering optical modelling, corresponding to a wave front error of 250 nm RMS at the VISPhot, FGS1, FGS2 and NIRSpec focal planes, and 280 nm RMS at the AIRS focal planes. The differences are negligible and it can be noted from the figure that an aberrated PSF behaves slightly better than a diffraction limited PSF, despite the former requires a larger number of pixel in each photometric or spectral bin aperture. However, because the aberrated PSF dilutes the signal more, pixels take longer to saturate, there are fewer exposure in a given observing time, hence read noise has overall a smaller impact.

ArielRad uncertainties estimates support the the science community to optimise the science of \ARIEL. For instance \citet{Changeat2019} used ArielRad estimates to investigate the capability of \ARIEL\ in  retrieving pressure-dependent chemical profiles from predictions of observed atmospheric spectra. Science analyses using ArielRad performance  estimates are ongoing, include aspects related to atmospheric retrieval, phase curve detection, transit timing variations, etc.,  and will be reported in \ARIEL\ phase B study report later in 2020 ahead of Mission Adoption: the \ARIEL\  "Red Book".

\section{Conclusions}

In this work we have discussed the algorithmic implementation of ArielRad, the \ARIEL\  radiometric simulator, used for the optimisation of the payload design and to evaluate the science performance of the ESA M4 space mission. ArielRad accounts for all relevant sources of uncertainties on the detection of exoplanetary atmospheres with \ARIEL, that are: photon noise (of astrophysical origin and from the instrument self emission), detector noise and electronic noise, and jitter noise. All other potential systematic of instrument origin are expected to be made negligible by a careful instrument design as detailed by \citet{Tinetti2018} and further discussed in the \ARIEL\ Yellow Book. 

ArielRad has been extensively validated against two alternative models, ExoSim and AERM, always showing excellent agreement at a few percent level across all \ARIEL\ bands. 


\ARIEL\ will perform the first statistical survey of the atmospheres of  a large and diverse sample, observing about 1000 exoplanets during its life time, as discussed in \citet{Edwards_2019} with the help of ArielRad. For all exoplanet observations, the photon noise of their host stars is the dominant source of uncertainties, as revealed by an ArielRad assessment.    

ArielRad performance estimates, in the form of noise vs wavelength achieved on a given set of exoplanetary targets, are a product distributed upon request. As the phase-B continues, the payload design is optimised in an iterative process that aims at building the most performant space mission within the envelope provided. An advancement in payload design does not however imply  the need to modify the algorithmic implementation of ArielRad thanks to its parametric description of the payload model. As a consequence, the simulator itself is planned to be released at the freeze-out of the payload design, to occur after mission adoption.

\begin{acknowledgements}
	We thank G. Pilbratt (ESA-ESTEC) for comments that greatly improved the development of ArielRad. 
\end{acknowledgements}

\section*{Appendices}

\appendix{Acronyms and symbols} 
\begin{table}[h]
	\caption{Symbols used in the main text ordered by comparison}
	\begin{tabular}{@{}ll@{}} 
		\hline\noalign{\smallskip}
		Symbol & meaning\\ 
		\noalign{\smallskip}\hline\noalign{\smallskip}
		$T_{14}$     &   time between first and fourth contact in transit \\
		$S_{Ph}$     &   Phoenix Spectral Energy Density \\
		$S$     &   stellar Spectral Energy Density \\
		$R_\star$     &   Stellar radius \\
		$D$     &   stellar distance \\
		$I_{zodi}$     &   zodiacal Background Spectral Energy Density \\
		$BB$     &   black body Spectral Energy Density \\
		$I_{inst}$     &   payload instrument Spectral Energy Density \\
		$I_{Inner}$     &   detector box Spectral Energy Density \\
		$\Phi$     &   optical transmission \\
		$Y$     &   channel identifier \\
		$QE$     &   quantum efficiency \\
		$PRF$     &   point response function \\
		$H_{pix}$     &   pixel response function \\
		$x, y$     &   pixels in the detector \\
		$l, m$     &   pixel indices in $x$ and $y$ axes \\
		$S_{T,Y}$     &   target signal in channel \\
		$\lambda_s$ & wavelength at spectral bin center\\
		$A_{tel}$     &   telescope area \\
		$A_p$     &   elliptical aperture in pixel used for aperture photometry in photometric channels \\
		$A_s$     &   rectangular box aperture of the size of spectral bin used in spectrometers \\
		$R$     &   spectral resolution \\
		$\Omega$     &   field of view \\
		$\Delta_{pix}$     &   pixel size \\
		$f_{eff}$     &   telescope effective focal lenght \\
		$S_{zodi,Y}$     &   zodiacal signal in channel \\
		$R_0$     &   spectral resolution at $\lambda=3.90 \mu m$ \\
		$L$     &   slith width expressed in number of pixel on the focal plane\\
	\noalign{\smallskip}\hline	
	\end{tabular}
	
	\label{aba:tbl1}
\end{table}

\begin{table}[h]
	\caption{Table~\ref{aba:tbl1} ({\it
			continued})}
	\begin{tabular}{@{}ll@{}} 
		\hline\noalign{\smallskip}
		Symbol & meaning\\ 
		\noalign{\smallskip}\hline\noalign{\smallskip}
		
		$S_{Inner}$     &   detector box signal in spectroscopic channels \\
		$S_{Y}$     &   total incoming signal in photometric or spectroscopic channels \\
		$S_{instr,Y}$     &   instrument signal in photometric or spectroscopic channels  \\
		$VAR()$     &   total variance  \\
		$VAR_P()$     &   Poisson noise contribution to the variance  \\
		$VAR_D()$     &   Detector noise contribution to the variance  \\
		$VAR_J()$     &   Jitter noise contribution to the variance  \\
		$\Delta t_{int}$     &  integration time expressed in hours\\
		$\chi_Y$      & margin added to Poisson noise\\
		$N_{pix}$     &   number of pixel included in spectral bin  \\
		$I_{dark}$     &   detector dark current  \\
		$n$     &   number of Non Destructive Reads  \\
		$T_{sat}$     &   detector saturation time  \\
		$WD$     &   pixel well depth  \\
		$f_{WD}$     &   coefficient describing the fraction of pixel well depth\\
		$\sigma_{rd}$ & pixel read noise\\
		$CR$ & contrast ratio \\
		$S_{OOT}$ & flux observed out-of-transit \\
		$S_{IT}$ & flux observed in-transit \\
		$S_P$ & flux from the planet \\
		$S_\star$ & flux from the star \\
		$\Delta z$ & atmosphere height \\
		$R_p$     &   planetary radius \\
		$S_{em}$ & flux from the planet thermal emission\\
		$A_l$ & fraction of starlight reflected by the planet\\
		$\alpha$ & geometrical albedo \\
		$a$ & semi-major axis in planet orbit \\
		$N_{min}$ & minimum noise considered in the channel in ESA radiometric model for \ARIEL\ \\ 
		\noalign{\smallskip}\hline
	\end{tabular}
\end{table}

\bibliographystyle{spbasic}
\bibliography{refs}

\end{document}